\documentclass[a4paper,twoside,psfig]{article}
\newcommand{\affil}[1]{$^{\rm #1}$}
\usepackage[authoryear]{natbib}
\bibpunct{(}{)}{;}{a}{}{,}
\usepackage{graphicx}
\usepackage{psfig}
\date{} %Please leave the date blank
\def\etal{{\it et al.~}}

\title{\large\bf\flushleft The Group Evolution Multiwavelength
Study (GEMS): the Sample and Datasets}
\author{\parbox{\textwidth}{\flushleft
\vspace{-0.5cm}
{\it Duncan A. Forbes\affil{A,F}, Trevor Ponman\affil{B}, Frazer Pearce\affil{C},
John Osmond\affil{D},  Virginia Kilborn\affil{A}, 
Sarah Brough\affil{A},  Somak Raychaudhury\affil{B}, Carole Mundell\affil{E}, Trevor
Miles\affil{B}, Katie Kern\affil{A}
}\\
\vspace{0.4cm}
{\small \affil{A}\,Centre for Astrophysics \& Supercomputing,
Mail H39, 
Swinburne University, Hawthorn VIC 3122, Australia}\\
{\small \affil{B}\,School of Physics and Astronomy, University of
Birmingham, Birmingham B15 2TT, UK}\\
{\small \affil{C}\,School of Physics and Astronomy, University of
Nottingham, Nottingham NG7 2RD, UK}\\
{\small \affil{D}\,Centre for Electronic Imaging, School of
Engineering and Design, Brunel University, Uxbridge
UB8 3PH, UK}\\
{\small \affil{E}\,Astrophysics Research Institute, Liverpool
John Moores University, Twelve Quays House, Birkenhead, CH41 1LD,
UK}\\
{\small \affil{F}\,Email: dforbes@swin.edu.au}}}
\begin{document}
\twocolumn[
\begin{changemargin}{.8cm}{.5cm}
\begin{minipage}{.9\textwidth}
\vspace{-1cm}
\maketitle
%
%
%%%%%%%%%%%%%     ABSTRACT    %%%%%%%%%%%%%
%Abstract of no more than 200 words here.
\small{\bf Abstract:}
Galaxy groups have been under-studied relative to their richer 
counterparts -- clusters. The Group Evolution Multiwavelength
Study (GEMS) aims to redress some the balance. Here we describe
the GEMS sample selection and resulting sample of 60 nearby 
(distance $<$ 130 Mpc) galaxy groups
and our multiwavelength dataset of X-ray, optical and HI
imaging. ROSAT X-ray images of each group are presented. 
GEMS also utilizes near-infrared imaging from the 2MASS survey and
optical spectra from the 6dFGS. These observational data are
complemented by mock group catalogues generated from the latest
$\Lambda$CDM simulations with gas physics included. Existing GEMS
publications are briefly
highlighted as are future publication plans. 

%%%%%%%%%%%%%     KEYWORDS    %%%%%%%%%%%%%
\medskip{\bf Keywords:} surveys: groups --- galaxies: evolution
--- galaxies: formation --- galaxies: groups 
% Please write all keywords in lower case. PASA uses the
% standard list of subject headings adopted by The Astrophysical Journal
% and available from http://www.journals.uchicago.edu/ApJ/keywords_text.html.
% Keywords are separated by em-dashes, i.e. ---

%%%%%%%%DO NOT EDIT%%%%%%%%%%%%
\medskip
\medskip
\end{minipage}
\end{changemargin}
]
\small
%%%%%%%%EDIT FROM HERE%%%%%%%%%%%%

\section{Introduction}

%Galaxies are thought to form in dark matter halos from overdensities in
%the early Universe.  After this `genetic imprint', however, their
%subsequent evolution is governed by the environment in which they
%live. 
The degree to which initial formation and environmental
influences determine galaxy properties today is similar to the `Nature
vs Nurture' debate.  Like the biological debate, this issue is
an ongoing and fundamental one for contemporary astrophysics.
In addition to the environment having a role in galaxy evolution,
the galaxies themselves also influence their local environment through expelled gas,
high-energy particles and radiation outflows. This may in turn
feedback into the subsequent evolution of galaxies within that
environment. 

Quantifying the importance of this complex interplay between galaxies and
their environment requires the detailed tracking of galaxy
properties over a broad range of environment and look-back
time. In terms of environment, there are three broad regimes
based on the spatial density of galaxies: rich clusters, groups
and the low-density `field'. At the highest densities, rich
clusters contain hundreds, even thousands, of galaxies within a
few cubic megaparsecs. In contrast, field galaxies are relatively
isolated galaxies whose nearest L$^{\ast}$ neighbours can be many megaparsecs
away. Galaxy groups represent an environment that is 
intermediate between these two extremes, containing a
few tens of galaxies within a cubic megaparsec. Despite the
meagre size of their populations, groups are vital to obtaining a
complete understanding of galaxy evolution and environmental
processes for the following reasons:\\

\noindent
\begin{itemize}
\item {\it Groups are 
the most common galaxy environment:} approximately 70\% of galaxies
in the local universe are located in groups (Tully 1987). 

\item {\it Groups are 
the building blocks of bigger structures:} in the hierarchical
clustering scenario, groups are the fundamental building blocks of galaxy
clusters. Hence the evolution of galaxies in groups is highly relevant to the early
evolution of rich cluster galaxies (e.g. Bekki 1999; Solanes
\etal 1999; Moss \& Whittle 2000). 

\item {\it Groups are 
the most favoured environment for galaxy mergers:} the 
best chance of observing galaxy mergers and interactions is in groups
because of the low relative velocities between galaxies; Mamon (2000) 
recently concluded that the merger rate is 100$\times$ higher in groups 
than in rich clusters. Groups are therefore the very best laboratory for
studying the merger process, which is known to be the most effective 
in transforming galaxies morphologically (e.g. Toomre \& Toomre
1972). 

\item {\it Groups contain a significant fraction  
of the hot gas in the universe:} the total amount of 
hot gas in groups is comparable
to that in  clusters 
and therefore provides 
a significant contributor to the baryonic component of the universe (Fukugita \etal
1998). Groups provide an unprecedented opportunity to study the nature 
and origin of this important mass component and the intimate link it
has with galaxies and their evolution.

\item {\it Star formation is suppressed on group scales:} large
surveys, such as the 2dFGRS (Lewis \etal 2002) have revealed that
star formation is suppressed at projected densities greater than
1--2 galaxies Mpc$^{-2}$. Such densities are associated with
groups. The physical mechanism for this `pre-processing' of
galaxies in groups is currently not well understood (e.g. Fujita
2003; Bower \& Balogh 2003). 

\item {\it A group is our local environment:} our Galaxy, the Milky Way,
lives in a small loose group containing several dozen galaxies. A comprehensive
understanding of galaxy groups in general will provide insight into our own
Local Group.
\end{itemize}

Clusters arise from high density pertubations. The 
early-type galaxies in clusters formed most of their stars 
at very early epochs, i.e. $z\ge 3$ (e.g. Bower, Lucey \& Ellis
1992; Kelson \etal 1997; Jorgensen \etal 1999; van Dokkum
\etal 2000; Terlevich \& Forbes 2002; Tanaka \etal 2005).
% and they
%tend to be located in the virialised (core) region of clusters.   
Galaxy groups, on the other hand, are seeded by much
lower density perturbations in a hierarchical Universe. They
range from early-collapsed systems (`fossil groups') to
overdensities only slightly greater than the `field'. Star
formation tends to be more extended in group galaxies (Terlevich
\& Forbes 2002).  
%, are still collapsing and virialisingtoday. 
Hence nearby groups offer 
the opportunity to study galaxy evolution and star formation
histories in systems of different dynamical
states.
In addition, their proximity means that they
are easily accessible with existing  telescopes at different
wavelengths. 

Previous studies of galaxy groups have included those focusing on
Hickson Compact Groups, e.g. the optical study by Hunsberger,
Charlton \& Zaritsky (1998), the X-ray study of Ponman \etal
(1996) and the HI survey by Verdes-Montenegro \etal (2001). On
the slightly scale, we have 
Poor (less than a dozen members), but relatively compact,
groups have also attracted some attention, e.g. Tovmassian,
Plionis \& Andernach (2004); Plionis, Basilakos \& Tovmassian
(2004).  Most
studies of large loose groups have been restricted to small samples,
perhaps due to the large areal coverage on the sky required. Wide
area optical surveys have been conducted by Ferguson \& Sandage
(1991) of 5 groups and Trentham \& Tully (2002) of 4
groups. Dedicated HI mapping of loose groups have been very 
few in number. Perhaps the largest is Maia, Willmer
\& da Costa (1998) who observed 73 (mostly loose) groups
but without uniform coverage. The HIPASS HI survey of the
southern sky contains around one hundred HI-detected groups
(Stevens 2005). X-ray studies have included similar numbers of loose
groups, e.g. Mulchaey \etal
(2000), Mahdavi \etal (2000) and Helsdon \etal (2001). 

However, most of these studies have tended to focus on a single
wavelength. A notable exception is the optical and X-ray study of groups
by Mulchaey \& Zabludoff (1998) and Zabludoff \& Mulchaey (1998), who
studied 3 compact and 9 loose groups. Their groups were selected optically
to be poor groups with less than 5 large galaxies.  Nine of their groups
were found to have extended X-ray emission, the remaining 3 were
undetected in the X-ray.  They found the 9 groups with intragroup X-ray
emission to be bound systems with many members (even the compact groups).
These bound groups appear to share a common, massive halo as
opposed to possessing individual dark matter halos. This may explain why the
galaxies in these groups have not yet all merged together.  They also found
bright elliptical galaxies at the centres of these groups, similar to
those seen in clusters, and fractions of early-type galaxies
approaching those
in clusters.  In contrast the 3 groups undetected in X-rays were found to
contain no early-type galaxies and have spiral galaxies at their centre.
They also found that, in terms of their optical (velocity dispersion) and
X-ray (X-ray luminosity and temperature) properties, groups are equivalent
to scaled-down clusters.

The physical nature of groups, evolutionary connection between
different types of groups and the effect of the group environment
on galaxies is not well understood. To fully understand any
astrophysical problem it is important to adopt a
multi-wavelength approach. This is particularly true for galaxy
groups for which we expect large energetic
and gas phase changes as a group evolves.

To overcome these limitations we created the Group Evolution
Multiwavelength Study (GEMS) which is a survey of 60 nearby loose and
compact galaxy groups. Here we describe the sample selection, the
existing and planned data sets, and an overview of group properties.
The GEMS collaboration involves researchers predominately from
Swinburne University, Australia and Birmingham University, UK,
but includes others from many astronomical institutions
world-wide. Further details can be obtained from our public web
pages at http://www.sr.bham.ac.uk/gems/.

\section{Sample Selection}

The GEMS sample is based on optically-defined groups which have
deep ROSAT X-ray data available. The presence of hot intragroup
gas indicates that the group is real (rather than merely 
a projection on the sky and that it is a bound system). The X-ray luminosity  
gives a measure of the virialisation
of the group's potential well and hence its evolutionary state.

The groups of galaxies were initially selected from 
a master group catalogue consisting of 4,320 optically-defined groups (e.g. Garcia
1993; Tully 1987).  
This master group list was compared
to the ROSAT satellite pointed observations, and all groups with a ROSAT
PSPC pointing within 20 arcmin of the group position were
extracted. We further required that 
the ROSAT observation must be greater than 10,000 s
(sufficient for detailed X--ray analysis), and that the recession
velocity of the group must lie between 1000 km\,s$^{-1} < v_{{\rm
group}} <$ 3000 km\,s$^{-1}$. This left a list of groups which should
neither be too close so as to fill the ROSAT field of view, nor
too distant for detailed analysis. After excluding a small
number of groups that are likely subclumps of the Hydra and Virgo
clusters, this left 45 groups. To this we added 13 groups 
from Helsdon \& Ponman (2000), and two Hickson
Compact Groups (HCG 4 and HCG 40) all with ROSAT PSPC imaging. 
Further details can be found in Osmond \& Ponman (2004). 

So although not a statistically-complete sample, the 60 groups
have a large range in properties such as X-ray luminosity,
dynamical state, velocity dispersion, richness and constituent galaxy types,
making them a representative sample of nearby galaxy groups and the
largest multiwavelength sample to date. They cover the whole
sky. The sample includes both loose and compact groups, thus we
are in a position to test possible evolutionary connections
between the two types of groups.

The distribution of the GEMS groups on the sky, in galactic
coordinates, is shown in Fig. 1. As can be seen in the figure, the GEMS groups
were chosen to lie at $|l| > 15^{o}$ from the Galactic plane. 

\begin{figure}[h]
\begin{center}
\includegraphics[scale=0.3, angle=-90]{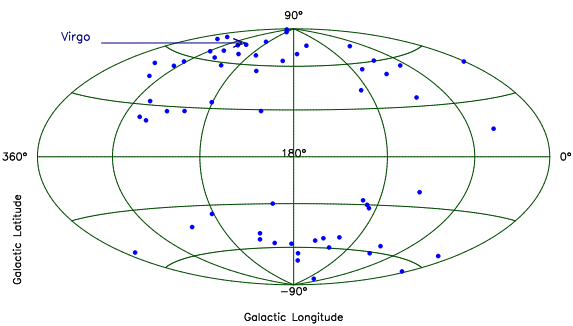}
\caption{
Distribution of the 60 GEMS groups on the sky in Galactic coordinates
shown in an equal-area Aitoff projection. Lines of equal latitude 
and longitude are marked at intervals of 30 and 60 deg respectively.
The position of the Virgo cluster
is marked.
}\label{figexample}
\end{center}
\end{figure}

\begin{figure}[h]
\begin{center}
\includegraphics[scale=0.3, angle=-90]{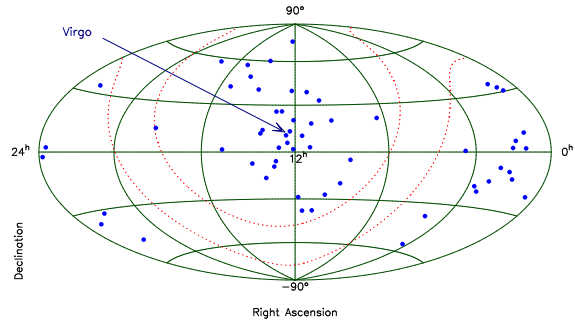}
\caption{
Distribution of the 60 GEMS groups on the sky in Equatorial coordinates
shown in an equal-area Aitoff projection. Lines of equal RA and 
Dec.  are marked at intervals of 60 and 30 deg respectively.
The survey avoids groups with Galactic Latitude $|b|<15^\circ$: these
limits are shown as dotted lines. 
The position of the Virgo cluster
is marked.
}\label{figexample}
\end{center}
\end{figure}

\section{A Multiwavelength Data Set}

GEMS uses public catalogues/archive data as well as new dedicated
observations. The main datasets are:\\

\noindent
$\bullet$ {\bf X-ray imaging.} All 60
groups have deep ($>$ 10,000s) ROSAT PSPC images available. The
ROSAT data, overlaid on DSS images, are shown in Figures
7--11. After taking into point sources, 37 groups reveal extended
X-ray emission associated with an intragroup medium (called G),
15 groups have non extended X-ray emission that we associate with
the galaxy halo (H) and 8 groups have X-ray
emission that is $<3\sigma$ times the background and are
classified as undetected (U). This diversity of X-ray properties
can be seen in Figures 6--10. For further details
see Osmond \& Ponman (2004). The ROSAT data are
complemented by ever increasing coverage from the XMM-Newton and
Chandra X-ray satellites (see Table 1), which have increased
sensitivity and/or spatial resolution.  \\

\noindent
$\bullet$ {\bf HI Mapping.} We have carried out wide area (5.5
$\times$ 5.5 sq. degrees) HI mapping with the Parkes radio telescope for 16
GEMS groups (see Table 1). Thus we sample out to radii of a few
Mpc or several virial radii. The
telescope was scanned across each group in a grid pattern for a
total of 16--20 hours. The resulting velocity resolution is
$\sim$2 km s$^{-1}$ (about 10$\times$ better than HIPASS) and a mass
limit in the range 4--10 $\times$ 10$^{8}$M$_{\odot}$ (about
2$\times$ better than HIPASS). The bandwidth is 8 MHz with 1024
spectral channels. The final beam size is 15.5 
arcmins. 
%Sources are found using a 3D
%visualisation tool developed at Swinburne University by
%P. Bourke. This 
Source selection is quantified using fake sources injected into
the datacubes and their recovery rate measured. 

A small number of Parkes sources have also been observed on the
Australia Telescope Compact Array. These are generally
sources that are `confused' in the large Parkes beam
or new group members (see Kilborn \etal 2005b for details).\\

\noindent
$\bullet$ {\bf Optical imaging.} Wide field imaging cameras have been
used to image the central 0.5 sq. degrees of 29 groups in
B,R,I filters. The data come from the 2.5m INT, Canary Islands
(17 groups), the 2.2m ESO/MPI telescope, Chile (8 groups) and the 3.9m AAT,
Australia (4 groups). Seeing was around 1 arcsec. 
Galaxy selection was carried out using the Sextractor
routine. The data reach to absolute magnitudes (assuming group
membership) of M$_B$ $\sim$ --13. For further details see Miles
\etal (2004). 

%We have begun to image some individual GEMS galaxies in the rest
%frame H$\alpha$ to study the global star formation rate and its 
%spatial distribution. An additional source of H$\alpha$ imaging
%is the SINGG survey (e.g. Meurer 2003). \\

\noindent
$\bullet$ {\bf Near-infrared imaging.} Image parameters and J,H,K magnitudes
are available for all GEMS group galaxies with K $<$ 13.1 from
the 2MASS survey (Jarrett \etal 2000). These near-infrared
magnitudes provide a good (photometric) tracer of galaxy mass,
and are used in GEMS to calculate luminosity-weighted group
centroid positions and analyse group properties 
(see Brough \etal 2005 for details).\\

\noindent
$\bullet$ {\bf Recession velocities.} Confirmed group membership
requires a recession velocity from either optical spectra or HI
observations. We complement the existing databases (such as NED
and Hyperleda) with our own new HI velocities and those from the
6dFGS (Jones \etal 2004) for some southern groups. More
velocities will become available after the third and final 6dFGS
data release. The SDSS survey will confirm group membership for
many northern groups. Such velocities are being used to derive
group properties such as velocity dispersions and hence probe
group dynamics (e.g. Brough \etal 2005) and can be used to test
whether HCGs are merely the dense cores of larger loose groups.\\

\section{Mock Catalogues}

Sample selection is one aspect of the study of galaxy groups that has
significantly hindered progress. A large, well-selected sample has
simply been lacking. This is partly due to the difficulty of defining
galaxy groups in an objective way from galaxy positions alone (see Eke
\etal 2004) combined with the difficulty of detecting X-ray emission 
from groups of $\le 10^{13}h^{-1}{\rm M}_{\odot}$ (e.g. 
Osmond \& Ponman 2004). To circumvent these problems and test our group selection
procedure we use modern simulations. 

\begin{figure}[h]
\begin{center}
\caption{Velocity dispersions from mock catalogues. The plot shows the
measured 3D$/\sqrt{3}$ (black points) and 1D (red points, in 3
orthogonal directions for each group/cluster) velocity dispersion
vs the true dark matter velocity dispersion of the halo for all
galaxies ${\rm M}_{B} < -16$ within the virial radius of
group-sized (M$ \ge 10^{13}h^{-1}{\rm M}_{\odot}$) and larger
halos. The solid blue line shows the one-to-one
relation. The plot contains
data for over 50,000 halos. At low velocity dispersion (small
groups), the 1D line-of-sight measurement becomes a less reliable
indicator of the true group velocity dispersion and the systems
themselves may not be true groups. [See attached jpg file.]  
}
\end{center}
\end{figure}

\begin{figure}[h]
\begin{center}
\includegraphics[scale=0.4, angle=0]{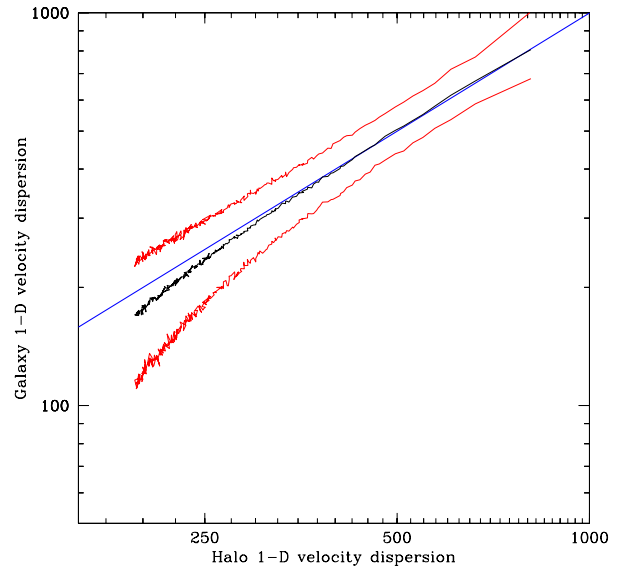}
\caption{The median velocity dispersions from the mock catalogues. Data
from Fig. 3 is reproduced as the black and red lines, which show the
mean, 10 percentile and 90 percentile for bins of 200 halos of the
specified dark matter velocity dispersion. For small haloes there is
a slight tendency for the galaxies to have a lower than expected
velocity dispersion. The solid blue line shows the one-to-one
relation. 
%The blue line shows the result of binning in sets
%of equivalent galaxy velocity dispersion, closer to the observational measure.
%if we measure a 1D velocity dispersion, what is the halo velocity
%dispersion likely to be ? For large haloes the measure velocity
%dispersion tends, on average, to overestimate the true velocity
%dispersion by around 6\%.  For group sized haloes a second effect appears; we are
%approaching the mass threshold of the model, so haloes must be
%{\it at
%least} of the minimum size. 
The plot shows that for low velocity dispersion groups
%should be treated
%with caution: such low velocity dispersions occur naturally in model
%samples and 
the dispersion is not a good indicator of the true group size or mass.
}
\end{center}
\end{figure}

\subsection{Dark matter and stellar content}

Our mock Universe is based on the $500h^{-1}{\rm Mpc}$ volume of
the Millennium simulation (Springel \etal 2005). With over
$10^{10}$ particles each of mass $8.6{\times}10^8h^{-1}{\rm
M}_{\odot}$ this model resolves halos containing typical galaxy
groups with over 10,000 particles.  There are over 50,000 halos
of group or larger size within the simulation volume, allowing
detailed statistics of the halo properties to be extracted.

%\subsection{Galaxy catalogues}

Within the framework of the  Millennium simulation we have access to
the mock catalogues of both Croton \etal (2005) and Bower \etal
(2005). Both these mock catalogues include AGN feedback and
successfully reproduce the break in the galaxy luminosity
function.

\begin{figure}[h]
\begin{center}
\caption{The number of galaxies within each halo. The black points
show the number of ${\rm M}_B<-16.32$ simulated 
galaxies within a cylinder of
radius $r_{500}$ as a function of the 1-D velocity dispersion of the
halo in the catalogues of Croton \etal (2005). 
The red points with error bars are the GEMS group sample of
Osmond \& Ponman (2004), also limited at ${\rm M}_B<-16.32$.
[See attached jpg file.]
}
\label{figngsig}
\end{center}
\end{figure}

\begin{figure}[h]
\begin{center}
\caption{The galaxy density within each halo. The black points 
show the measured galaxy
density for ${\rm M}_B<-16.32$ simulated 
galaxies, where we have taken the
volume of a sphere of radius $r_{500}$ but the number of galaxies
counted within a cylinder of this radius (to match the observational
method). We see increased scatter in galaxy density at 
low velocity dispersion groups, 
which are only partially resolved in our simulations. 
%Galaxy densities below 100 ${\rm M}_B<-16.32$ per
%$h^{-1} Mpc^{3}$ only start to appear in 
The red squares with errorbars are the GEMS groups of Osmond
\& Ponman (2004); 
open squares indicating the groups with unreliable sizes. 
[See attached jpg file.]
}
\label{figrhogal}
\end{center}
\end{figure}

\subsection{Gas properties}

The $500h^{-1}{\rm Mpc}$ volume of the Millennium simulation has also
been modelled at lower resolution but including
hydrodynamics. Although this lower resolution is insufficient for a full
treatment of the gaseous phase of galaxies it is
sufficient to resolve the hot halos and produce reliable X-ray
luminosities and gas temperatures. We have two runs, one without gas
cooling (which produces halos that are far too bright in X-rays) and a
run with preheating that has been tuned to reproduce both the observed
X-ray luminosity--temperature relation and the mass--temperature
relation. These models accurately reproduce the halo distribution of
the higher resolution Millennium simulation. 

With this set of models we can inter-compare galaxy, dark matter and gas
properties for a volume limited sample of $>50,000$ halos. We
can also examine the evolution of any desired halo or set of halos, as both
the dark matter and galaxy catalogues have 64 output redshifts, a
number increased further still to 160 in the gas models.

As an example of what is available, we have extracted the 1D dark
matter velocity dispersion within the virial radius of all the haloes
with ${\rm M} \ge 10^{13}h^{-1}{\rm M}_{\odot}$. In Fig. 3 we compare
this to the 3D$/\sqrt{3}$ (black points) and 1D (red points) velocity dispersion
for all galaxies with $ {\rm M}_B < -16$ within these halos. At low
velocity dispersion (small groups), the 1D line-of-sight measurement
becomes a less reliable indicator of the true group velocity
dispersion.

Figure~4 displays the same information but this time only the median
, 10th and 90th percentile values are shown. 
For small haloes there is
a slight tendency for the galaxies to have a lower than expected
velocity dispersion, due to the fact that the semi-analytic galaxy is
always the most-bound particle of its parent halo. No correction
is made for the 
small numbers of galaxies or the fact that the central galaxy is
assumed to be at rest with respect to the group.
%The blue line is an attempt to recover something 
%closer to the {\it observed} measure, as we cannot
%easily measure the dark matter velocity dispersion. In this case we
%have ranked the groups in order of the velocity dispersion derived
%from the galaxies and again plotted bins of 200 groups. 
Due to the
fact that we have over 50,000 clusters and 150,000 independent
measures of the velocity dispersion our sample contains several
hundred groups with derived velocity dispersions below $100{\rm
km/s}$. If we solely used the galaxy velocity dispersion to derive the
size of these groups we would arrive at far too small a number. 
The sharp drop seen in Fig. 3 is mainly driven because we have deliberately chosen
to cut off our sample at $10^{13}h^{-1}M_{\odot}$ so that there are
suddenly no smaller groups in our sample. Something like this effect
also occurs observationally: below a certain size, groups are not
identified because they only contain a single galaxy. Slightly 
above this threshold the recovered velocity dispersion is, on average,
6\% too high:
this is simply because there are more smaller groups than larger ones,
so more groups scatter up than down. 
In summary, velocity dispersion measures for low
mass groups become increasingly unreliable with a tendency to 
underestimate the group velocity dispersion (and hence
underestimate group mass).
%Low velocity dispersion groups should therefore
%be treated with caution, i.e. low velocity dispersions occur naturally in model
%samples but are not a good indicator of the true group size or mass.

\subsection{Galaxy counts}

The number of galaxies above a specified brightness (in this case
${\rm M}_B<-16.32$)  is well correlated with the halo velocity
dispersion (Fig.~5). The GEMS group sample of Osmond \&
Ponman (2004) spans the model results and extends them to lower
velocity dispersions. Many of the lower mass GEMS groups may not
be virialised systems.  
The galaxy density (for galaxies brighter than 
${\rm M}_B<-16.32$) is stable at a value of around 45 galaxies per 
$h^{-1}Mpc^{3}$ 
although the scatter increases dramatically as the velocity dispersion
of the halo drops (Fig. 6). Again the Osmond \& Ponman groups span the model
range, with no trend for a large systematic increase in galaxy density
as the velocity dispersion is lowered; such a trend would be a clear
indication that the size of the groups (and hence masses) had been
systematically underestimated. This is not a problem for the GEMS
groups, as discussed in Osmond \& Ponman (2004), section 6.

\section{Group Properties}
%Please see the PASA Style Guide for help with correct layout for your manuscript.
%Examples of tables and figures are given below.
We summarise some basic properties of the 60 GEMS groups in Table
1. The group name is generally the brightest central galaxy or
Hickson Compact Group number. The  
position corresponds to the centrally located galaxy or that
listed in the optical group catalog (see Osmond \& Ponman (2004)). The distance is
calculated from a Virgo-infall corrected
velocity and assuming H$_0$ = 70 km/s/Mpc (this Hubble constant
is used in all GEMS papers). The r$_{500}$ radius
is a measure of the extent of each group. It is defined as the
radius for which the density is 500$\times$
the critical density of the Universe. For groups, this quantity is more
reliably obtained than the commonly used r$_{200}$ or virial
radius (r$_{500}$ $\sim$ 2/3 r$_{200}$) in clusters. It is
calculated based on the simulations of Evrard et al. (1996) 
using the measured X-ray temperature, T$_X$ (for those groups without X-ray
temperatures, it is calculated from the summed B-band luminosity
(L$_B$) of all
galaxies in the group). The r$_{500}$ values listed here have
been updated from Osmond \& Ponman (2004) so that for groups with
galaxy halo emission the radius is now calculated based on L$_B$
(rather than T$_X$).  
The measured X-ray luminosity and upper limits are from
Osmond \& Ponman (2004). Whether this X-ray emission comes from  
intragroup emission (G), 
galaxy halo emission (H) or is undetected in
X-rays (U) is listed. The current existing datasets for each group are given by
the codes: O = optical imaging, P = Parkes HI mapping, C = Chandra
X-ray imaging, X = XMM X-ray imaging.

\begin{figure*}
\begin{center}
\resizebox{40pc}{!}{
\begin{tabular}{ccc}
\end{tabular}
}
\end{center}
\caption{The GEMS groups. ROSAT X-ray contours overlaid on
optical DSS images. The dashed line represents the `cut-off
radius' at which the X-ray emission reaches the background level.
The small yellow circles indicate point sources. Large yellow
circles indicate that point sources dominate that region.  
[See attached jpg file.]
}
\label{fig:xray1}
\end{figure*}

\begin{figure*}
\begin{center}
\resizebox{40pc}{!}{
\begin{tabular}{ccc}
\end{tabular}
}
\end{center}
\caption{The GEMS groups. ROSAT X-ray contours overlaid on optical DSS images. 
The dashed line represents the `cut-off
radius' at which the X-ray emission reaches the background level.
The small yellow circles indicate point sources. Large yellow
circles indicate that point sources dominate that region.
[See attached jpg file.]
}
\label{fig:xray1}
\end{figure*}

\begin{figure*}
\begin{center}
\resizebox{40pc}{!}{
\begin{tabular}{ccc}
\end{tabular}
}
\end{center}
\caption{The GEMS groups. ROSAT X-ray contours overlaid on optical DSS images. 
The dashed line represents the `cut-off
radius' at which the X-ray emission reaches the background level.
The small yellow circles indicate point sources. Large yellow
circles indicate that point sources dominate that region.
[See attached jpg file.]
}
\label{fig:xray1}
\end{figure*}

\begin{figure*}
\begin{center}
\resizebox{40pc}{!}{
\begin{tabular}{ccc}
\end{tabular}
}
\end{center}
\caption{The GEMS groups. ROSAT X-ray contours overlaid on optical DSS images. 
The dashed line represents the `cut-off
radius' at which the X-ray emission reaches the background level.
The small yellow circles indicate point sources. Large yellow
circles indicate that point sources dominate that region.
[See attached jpg file.]
}
\label{fig:xray1}
\end{figure*}

\begin{figure*}
\begin{center}
\resizebox{40pc}{!}{
\begin{tabular}{ccc}
\end{tabular}
}
\end{center}
\caption{The GEMS groups. ROSAT X-ray contours overlaid on optical DSS images. 
The dashed line represents the `cut-off
radius' at which the X-ray emission reaches the background level.
The small yellow circles indicate point sources. Large yellow
circles indicate that point sources dominate that region.
[See attached jpg file.]
}
\label{fig:xray1}
\end{figure*}

%%Format tables as in the following example
%\begin{table}[h]
%\begin{center}
%\caption{Example Table}\label{tableexample}
%\begin{tabular}{lcc}
%\hline Group & Column 2 & Column 3 \\
%\hline Table Content$^a$ \\
%\hline
%\end{tabular}
%\medskip\\
%$^a$Group name, ***.\\
%\end{center}
%\end{table}

\begin{table*}
\begin{center}
\caption{GEMS Group Properties}\label{tableexample}
\scriptsize
\begin{tabular}{lccccccc}
\hline

%%%%%%%%%%%%%%%%%%%%%%%%%%%%%%%%%%%%%%%%%%%%%%%%%%%%%%%%%%%%%%%%%%%%%%%%%%%%%%%%%

Group     &  R.A.         &  Dec.     & Distance & r$_{500}$ &
L$_X$ & X-ray & Datasets\\
Name      &  (J2000)     &  (J2000)   & (Mpc)  & (Mpc) & (erg/s)
& class & \\

\hline
     
HCG 4     &  00 34 13.8  &  -21 26 21  & 115 &  0.36$^{\ast}$  &  41.48 $\pm$ 0.19  &  G & C,X\\
NGC 315   &  00 57 48.9  &  +30 21 09  & 72  &  0.55  &  41.21 $\pm$ 0.10  &  G & C\\
NGC 383   &  01 07 24.9  &  +32 24 45  & 73  &  0.69  &  43.07 $\pm$ 0.01  &  G & C\\
NGC 524   &  01 24 47.8  &  +09 32 19  & 35  &  0.42  &  41.05 $\pm$ 0.05  &  H & P,O\\
NGC 533   &  01 25 31.3  &  +01 45 33  & 76  &  0.58  &  42.67 $\pm$ 0.03  &  G & C,X\\
HCG 10    &  01 25 40.4  &  +34 42 48  & 68  &  0.24  &  41.70 $\pm$ 0.14  &  G & O\\
NGC 720   &  01 53 00.4  &  -13 44 18  & 23  &  0.40  &  41.20 $\pm$ 0.02  &  G & P,O,C,X\\
NGC 741   &  01 56 21.0  &  +05 37 44  & 79  &  0.62  &  42.44 $\pm$ 0.06  &  G & C\\
HCG 15    &  02 07 37.5  &  +02 10 50  & 95  &  0.54  &  42.12 $\pm$ 0.05  &  G & X\\
HCG 16    &  02 09 24.7  &  -10 08 11  & 57  &  0.32  &  41.30 $\pm$ 0.11  &  G & C,X\\
NGC 1052  &  02 41 04.8  &  -08 15 21  & 20  &  0.24  &  40.08 $\pm$ 0.15  &  H & P,O,C,X\\
HCG 22    &  03 03 31.0  &  -15 41 10  & 39  &  0.29  &  40.68 $\pm$ 0.13  &  G & O\\
NGC 1332  &  03 26 17.1  &  -21 20 05  & 23  &  0.25  &  40.81 $\pm$ 0.02  &  H & P,O,C\\
NGC 1407  &  03 40 11.8  &  -18 34 48  & 26  &  0.57  &  41.69 $\pm$ 0.02  &  G & P,O,C\\
NGC 1566  &  04 20 00.6  &  -54 56 17  & 21  &  0.53  &  40.41 $\pm$ 0.05  &  H & P,O\\
NGC 1587  &  04 30 39.9  &  +00 39 43  & 55  &  0.55  &  41.18 $\pm$ 0.09  &  G & C\\
NGC 1808  &  05 07 42.3  &  -37 30 46  & 17  &  0.32$^{\ast}$  &  $<$40.10          &  U & P,O,C,X\\
NGC 2563  &  08 20 35.7  &  +21 04 04  & 73  &  0.57  &  42.50 $\pm$ 0.03  &  G & O,X\\
HCG 40    &  09 38 54.5  &  -04 51 07  & 102 &  0.45$^{\ast}$  &  $<$41.04          &  U & --\\
HCG 42    &  10 00 14.2  &  -19 38 03  & 64  &  0.48  &  41.99 $\pm$ 0.02  &  G & C,X\\
NGC 3227  &  10 23 30.6  &  +19 51 54  & 27  &  0.35$^{\ast}$  &  41.23 $\pm$ 0.05  &  H & O,C,X\\
HCG 48    &  10 37 49.5  &  -27 07 18  & 41  &  0.23$^{\ast}$  &  41.09 $\pm$ 0.04  &  G & --\\
NGC 3396  &  10 49 55.2  &  +32 59 27  & 31  &  0.36  &  40.53 $\pm$ 0.08  &  H & O,C,X\\
NGC 3557  &  11 09 57.4  &  -37 32 17  & 39  &  0.27  &  42.04 $\pm$ 0.04  &  G & P,O,C\\
NGC 3607  &  11 16 54.7  &  +18 03 06  & 23  &  0.33  &  41.05 $\pm$ 0.05  &  G & O,X\\
NGC 3640  &  11 21 06.9  &  +03 14 06  & 29  &  0.35$^{\ast}$  &  $<$40.37          &  U & O\\
NGC 3665  &  11 24 43.4  &  +38 45 44  & 37  &  0.38  &  41.11 $\pm$ 0.08  &  G & O,C,X\\
NGC 3783  &  11 39 01.8  &  -37 44 19  & 36  &  0.25$^{\ast}$  &  40.76 $\pm$ 0.11  &  G & P,O,C,X\\
HCG 58    &  11 42 23.7  &  +10 15 51  & 98  &  0.51$^{\ast}$  &  $<$41.33          &  U & --\\
NGC 3923  &  11 51 02.1  &  -28 48 23  & 22  &  0.36  &  40.98 $\pm$ 0.02  &  H & P,O,C,X\\
NGC 4065  &  12 04 06.2  &  +20 14 06  & 106 &  0.62  &  42.64 $\pm$ 0.05  &  G & X\\
NGC 4073  &  12 04 27.0  &  +01 53 48  & 96  &  0.69  &  43.41 $\pm$ 0.02  &  G & C,X\\
NGC 4151  &  12 10 32.6  &  +39 24 21  & 23  &  0.29$^{\ast}$  &  $<$40.20          &  U & O,C,X\\
NGC 4193  &  12 13 53.6  &  +13 10 22  & 39  &  0.39$^{\ast}$  &  40.63 $\pm$ 0.08  &  H & --\\
NGC 4261  &  12 19 23.2  &  +05 49 31  & 41  &  0.64  &  41.92 $\pm$ 0.03  &  G & O,C,X\\
NGC 4325  &  12 23 06.7  &  +10 37 16  & 117 &  0.51  &  43.15 $\pm$ 0.01  &  G & C,X\\
NGC 4589  &  12 21 45.0  &  +75 18 43  & 29  &  0.43  &  41.61 $\pm$ 0.05  &  G & X\\
NGC 4565  &  12 36 20.8  &  +25 59 16  & 27  &  0.49  &  40.44 $\pm$ 0.14  &  H & C,X\\
NGC 4636  &  12 42 50.4  &  +02 41 24  & 10  &  0.51  &  41.49 $\pm$ 0.02  &  G & P,O,C,X\\
NGC 4697  &  12 48 35.7  &  -05 48 03  & 20  &  0.39  &  41.01 $\pm$ 0.02  &  H & O,C,X\\
NGC 4725  &  12 50 26.6  &  +25 30 06  & 25  &  0.43  &  40.63 $\pm$ 0.06  &  H & O,C,X\\
HCG 62    &  12 53 05.8  &  -09 12 16  & 74  &  0.67  &  43.14 $\pm$ 0.04  &  G & C,X\\
NGC 5044  &  13 15 24.0  &  -16 23 06  & 33  &  0.62  &  43.01 $\pm$ 0.01  &  G & P,O,C,X\\
NGC 5129  &  13 24 10.0  &  +13 58 36  & 108 &  0.51  &  42.33 $\pm$ 0.04  &  G & X\\
NGC 5171  &  13 29 21.6  &  +11 44 07  & 107 &  0.58  &  42.38 $\pm$ 0.06  &  G & C,X\\
HCG 67    &  13 49 11.4  &  -07 13 28  & 115 &  0.46  &  42.02 $\pm$ 0.07  &  G & --\\
NGC 5322  &  13 49 15.5  &  +60 11 28  & 35  &  0.42  &  40.71 $\pm$ 0.10  &  H & O,X\\
HCG 68    &  13 53 26.7  &  +40 16 59  & 41  &  0.43  &  41.52 $\pm$ 0.04  &  G & X\\
NGC 5689  &  14 34 52.0  &  +48 39 36  & 38  &  0.26$^{\ast}$  &  $<$40.24          &  U & X\\
NGC 5846  &  15 06 29.2  &  +01 36 21  & 30  &  0.48  &  41.90 $\pm$ 0.02  &  G & O,C,X\\
NGC 5907  &  15 15 53.9  &  +56 19 46  & 17  &  0.24$^{\ast}$  &  39.69 $\pm$ 0.14  &  H & X\\
NGC 5930  &  15 26 07.9  &  +41 40 34  & 41  &  0.50  &  40.73 $\pm$ 0.07  &  H & --\\
NGC 6338  &  17 15 22.9  &  +57 24 40  & 127 &  0.88$^{\ast}$  &  43.51 $\pm$ 0.02  &  G & C\\
NGC 6574  &  18 12 00.7  &  +14 02 44  & 35  &  0.16$^{\ast}$  &  $<$40.81          &  U & \\
NGC 7144  &  21 52 42.9  &  -48 15 16  & 27  &  0.30  &  40.33 $\pm$ 0.13  &  H & P,O\\
HCG 90    &  22 02 08.4  &  -31 59 30  & 36  &  0.38  &  41.49 $\pm$ 0.05  &  G & P,O,C,X\\
HCG 92    &  22 35 58.4  &  +33 57 57  & 88  &  0.47  &  41.99 $\pm$ 0.04  &  G & C,X\\
IC  1459  &  22 57 10.6  &  -36 27 44  & 26  &  0.35  &  41.28 $\pm$ 0.04  &  G & P,O,C,X\\
NGC 7714  &  23 36 14.1  &  +02 09 19  & 39  &  0.22$^{\ast}$  &  $<$40.03          &  U & P,X\\
HCG 97    &  23 47 22.9  &  -02 18 02  & 92  &  0.51  &  42.37 $\pm$ 0.05  &  G & X\\

%%%%%%%%%%%%%%%%%%%%%%%%%%%%%%%%%%%%%%%%%%%%%%%%%%%%%%%%%%%%%%%%%%%%%%%%%%%%%%%%%

\hline
\end{tabular}
\medskip\\
$^a$Group name (generally the brightest central galaxy or HCG
number), 
position of centrally located galaxy or
optical group catalog, distance from Virgo-infall corrected
velocity and assuming H$_0$ = 70 km/s/Mpc, radius at 500$\times$
the critical density ($^{\ast}$ = uncertain radius), group X-ray class (G = intragroup emission, H =
galaxy halo emission, U = undetected in X-rays), O = optical, P =
Parkes, C = Chandra, X = XMM.\\
\end{center}

%%%%%%%%%%%%%%%%%%%%%%%%%%%%%%%%%%%%%%%%%%%%%%%%%%%%%%%%%%%%%%%%%%%%%%%%%%%%%%%%%
\end{table*}

\section{GEMS publications}

Since the first presentation at the European National Astronomy
meeting (McKay \etal 2002),
there have been several GEMS publications:\\

\noindent
$\bullet$ {\bf Khosroshahi \etal (2004).} An 
analysis of galaxy surface brightness profiles in 16 groups,
which found galaxies in X-ray luminous groups to be in general
larger and more luminous than their counterparts in low X-ray
groups. \\

\noindent
$\bullet$ {\bf Osmond \& Ponman (2004).} An analysis of
the X-ray properties of the entire sample. X-ray luminosities to
a uniform group scale radius are derived. More recent fits to the
L$_X$ scaling relations (Ponman \etal 2006) indicate that the
X-ray properties of groups are consistent with being 
scaled-down clusters. Various group optical
properties are found to be correlated to X-ray properties 
(e.g. the spiral fraction with IGM X-ray temperature). \\ 

\noindent
$\bullet$ {\bf McKay \etal (2004).} This paper reports the discovery
of two new galaxies, from Parkes HI mapping, in the NGC 1052 and
NGC 5044 groups. The two galaxies have HI masses of 0.5--1.0
$\times$ 10$^{9}$ M$_{\odot}$ and faint optical counterparts. \\

\noindent
$\bullet$ {\bf Miles \etal (2004).} Luminosity functions, in the B
and R bands, are presented for the central regions of 25
groups. A strong dip in the LF is seen around M$_B$ = --17 in the
low X-ray groups. This may be interpreted as a signature of
merging at intermediate galaxy mass scales in pre-virialised,
collapsing groups.\\

\noindent
$\bullet$ {\bf Kilborn \etal (2005a).} This paper presents wide area
Parkes HI mapping of the low X-ray emission group NGC 1566. Of the 13 HI
detected galaxies, two were not known previously to be group
members. Several galaxies appear be HI deficient; the cause of which
is probably tidal interactions rather than ram pressure stripping.\\

\noindent
$\bullet$ {\bf Brough \etal (2005).} The dynamics of the GEMS groups NGC 1407 and
NGC 1332 are studied, along with the Eridanus cloud. It is
suggested these groups form a bound `supergroup' of 7 $\times$
10$^{13}$ M$_{\odot}$ that is due to merge into a cluster.\\

\noindent
$\bullet$ {\bf Kilborn \etal (2005b).} This paper presents wide area
Parkes HI mapping of the NGC 3783 group, which has some evidence
for a weak IGM. We find one HI `cloud' (i.e. detected HI with no
optical counterpart to faint levels) of mass $\sim$ 10$^{9}$
M$_{\odot}$. This cloud is likely to have a tidal origin. \\

\section{Summary and Future Work}

Groups are the preferred environment of galaxies for more than
half of the age of the Universe. The GEMS project is an attempt to
better quantify and understand the processes operating in nearby galaxy
groups, building on the work of Zabludoff, Mulchaey and others
from the 1990s. Our sample of 60 groups covers a range of group
dynamical states. Our multiwavelength dataset and mock catalogues
allow us to define the properties of these groups and better
understand the
physical processes operating. Our existing, and future, GEMS
publications will provide a useful `local benchmark' for groups
found at moderate to high redshift in galaxy surveys such as the
2dFGRS, SDSS, DEEP2, 6dFGS. 

Future work includes measuring the K-band galaxy
luminosity function, a dynamical analysis of galaxy motions, high spatial
resolution HI follow-up of selected galaxies, examination of the
star formation rates via H$\alpha$ fluxes and the origin of brightest group
galaxies. 

\section*{Acknowledgments} %If needed

Thanks to ARC for funding aspects of this research and the telescope
support staff at various observatories who helped collect the
GEMS datasets.  
We thank the following for their help with aspects of GEMS:
D. Barnes, P. Goudfrooij,  S. Helsdon,  H. Khosroshahi,
B. Koribalski, N. McKay, 
P. Nulsen, R. Proctor.

%\end{multicols}


\begin{thebibliography}{}
% References are listed as in the following example, for more examples, please
% see the PASA Style Guide
%\bibitem[Smith, Jones, \& Brown(Year)Smith et al.]{example}Smith, A.~B., Jones,C.~D., Brown, E.~F. Year, Journal, Volume, Page
\bibitem[]{}Bekki, K. 1999 ApJL, 510, 15
\bibitem[]{}Bower, R., Lucey, J., Ellis, R., 1992, MNRAS, 254, 601
\bibitem[]{}Bower, R., Balogh, M., 2003,  RMxAC, 17, 220
\bibitem[]{}Bower, R., \etal 2005, astro-ph/0511338
\bibitem[]{}Brough \etal 2005, MNRAS, submitted
\bibitem[]{}Hunsberger, S., Charlton, J., Zaritsky, D., 1998,
ApJ, 505, 536
\bibitem[]{}Croton, \etal 2005, astro-ph/0508046
\bibitem[]{}Eke, V., \etal 2004, MNRAS, 348, 866
\bibitem[]{}Evrard, A., Metzler, C., Navarro, J., 1996, ApJ, 469,
494
\bibitem[]{}Ferguson, H., Sandage, A., 1991, AJ, 101, 765
\bibitem[]{}Fukugita, M., Hogan, C., Peebles, J., 1998, ApJ, 503,
518
\bibitem[]{}Fujita, Y., 2003, astro-ph/0311193
\bibitem[]{}Garcia, A., 1993, A\&AS, 100, 47
\bibitem[]{}Helsdon, S., Ponman, T., 2000, MNRAS, 319, 933
\bibitem[]{}Helsdon, S., Ponman, T., O'Sullivan, E., Forbes, D., 
2001, MNRAS, 325, 693
\bibitem[]{}Jarrett, T.,  Chester, T., Cutri, R., Schneider, S.,
Skrutskie, M., Huchra, J., 2000, AJ, 119, 2498
\bibitem[]{}Jones, D., \etal 2004, MNRAS, 355, 747
\bibitem[]{}Jorgensen, I., \etal 1999, MNRAS, 308, 833
\bibitem[]{}Kelson, D., \etal 1997, ApJL, 478, 13
\bibitem[]{}Kilborn \etal 2005a, MNRAS, 356, 77
\bibitem[]{}Kilborn \etal 2005b, MNRAS, submitted
\bibitem[]{}Khosroshahi \etal 2004, MNRAS, 349, 527
\bibitem[]{}Kilborn \etal 2005, MNRAS, 356, 77
\bibitem[]{}Lewis, I., \etal 2002, MNRAS, 334, 673
\bibitem[]{}Mahdavi, A., Bohringer, H., Geller, M., Ramella, M.,
2000, ApJ, 534, 114
\bibitem[]{}Maia M.A.G., Willmer C.N.A., da Costa L.N., 1998, AJ, 115, 49
\bibitem[]{}Mamon, G., 2000, In ``Dynamics of Galaxies: from the
Early Universe to the Present'', Ed. Francoise Combes, Gary
A. Mamon, and Vassilis Charmandaris, ASP Conference Series, p. 377
\bibitem[]{}McKay \etal  2004, MNRAS, 352, 1121
\bibitem[]{}Meurer, G., 2003, astro-ph/0311184
\bibitem[]{}Miles \etal  2004, MNRAS, 355, 785
\bibitem[]{}Moss, C. \& Whittle, M. 2000, MNRAS, 317, 667
\bibitem[]{}Mulchaey, J., 2000, ARAA, 38, 289
\bibitem[]{}Mulchaey, J., Zabludoff, A., 1998, ApJ, 496, 73
\bibitem[]{}Osmond, J., Ponman, T.,  2004, MNRAS, 350, 1511
\bibitem[]{}Plionis, M., Basilakos, S., Tovmassian, H., 2004,
MNRAS, 352, 1323
\bibitem[]{}Ponman, T., Bourner, P., Ebeling, H., Bohringer, H.,
1996, MNRAS, 283, 690
\bibitem[]{}Ponman \etal 2006, MNRAS, in prep.
\bibitem[]{}Solanes, J.M., Salvador-Sol\'e,
E. \& Gonz\'alez-Casado, G., 1999, A\&A, 343, 733
\bibitem[]{}Springel, V., \etal 2005, Nature, 435, 629
\bibitem[]{}Steven, J., 2005, PhD Thesis, University of Melbourne
\bibitem[]{}Tanaka, M., \etal 2005, MNRAS, in press
\bibitem[]{}Terlevich, A., Forbes, D., 2002, MNRAS, 330, 547
\bibitem[]{}Toomre, A., Toomre, J., 1972, ApJ, 178, 623 
\bibitem[]{}Tovmassian, H., Plionis, M., Andernach, H., 2004,
ApJL, 617, 111 
\bibitem[]{}Trentham, N., Tully, R., 2002, MNRAS, 335, 712 
\bibitem[]{}Tully, R. B., 1987, ApJ, 321, 280
\bibitem[]{}van Dokkum, P., \etal 2000, ApJ, 541, 95
\bibitem[]{}Verdes-Montenegro, L., Yun, M., Williams, B.,
Huchtmeier, W., Del Olmo, A., Perea, J., 2001, A\&A, 377, 812
\bibitem[]{}Zabludoff, A., Mulchaey, J., 1998, ApJ, 496, 39


\end{thebibliography}
\end{document}